# On the Non-Oxygenic Origins of Thylakoids

Luc Cornet[1*]

1 BCCM/ULC, InBioS–Molecular Diversity and Ecology of Cyanobacteria, University of Liège, 4000 Liège, Belgium
* Author to whom correspondence should be addressed.

## Abstract

**Thylakoid membranes are the site of oxygenic photosynthesis, one of the most important biochemical processes on earth. The ancestral state of these membranes is represented today in *Gloeobacterales*, where they are lacking and photosynthesis instead takes place in the cytoplasmic membrane. The evolutionary transition from this ancestral state to the modern thylakoid membranes was a major advantage, as it increased surface area devoted to the photosystems and thus photosynthetic efficiency. However, how this transition occurred remains a significant and surprisingly understudied biological question. With a highly synchronized process involving numerous assembly factors, the biogenesis of thylakoid membranes suggests the existence of intermediate evolutionary states during the emergence of this compartment. Here, I propose a non-oxygenic origin of thylakoid membranes, where these intermediate states were initially dedicated to alternative electron flows. This hypothesis addresses the paradox of cyanobacterial diversification in an euxinic environment, which is toxic to photosystem II.**

## Main text

Oxygenic photosynthesis (OxyP) is a biochemical process of high importance on earth, catalyzing the conversion of solar energy into carbohydrates by fixing inorganic carbon and releasing oxygen as a byproduct. Numerous recent evolutionary studies have significantly advanced our understanding of OxyP, including the origins of photosystems (PS) (Nishihara et al. 2024), PS evolution (Cardona 2018; Cardona et al. 2019; Oliver et al. 2021), and even dating its emergence (Fournier et al. 2021; Sánchez-Baracaldo et al. 2022). OxyP operates through a linear electron transfer (LET) chain that mainly occurs in the membranes of specialized structures called thylakoids. The selective advantage provided by thylakoid membranes (TM), in terms of number of PS per cell, has recently been proposed (Guéguen and Maréchal 2022) as a plausible explanation for earth's oxygenation during the Great Oxidation Event (GOE), 2.4 billion years ago (Bekker et al. 2004). However, despite extensive studies on the origins of OxyP and PS, the evolution of TM has been comparatively understudied. In most cyanobacteria, here referred to as Phycobacteria (Cavalier-Smith 2002), OxyP occurs in TM. Yet, the earliest-diverging group of cyanobacteria, the *Gloeobacterales*—as confirmed by multiple phylogenomic studies (Criscuolo and Gribaldo 2011; Shih et al. 2013; Soo et al. 2014; Uyeda et al. 2016)—lacks these membranes. Instead, OxyP takes place in their cytoplasmic membrane (CM) (Rippka et al. 1974; Guglielmi et al. 1981; Rexroth et al. 2011; Rahmatpour et al. 2021), which constitutes the ancestral state (**Figure 1A**). How the transition unfolded, from this ancestral *Gloeobacter*-like state, with a LET chain functioning in the CM, to the modern state, where all components are localized in TM, remains a biological enigma. Schematically, two distinct steps must be considered in this evolutionary process: first, the emergence of the cellular compartment, and second, the transfer of LET-related complexes from the CM to the TM.

TM consist of three key glycolipids, which are also present in the CM: monogalactosyldiacylglycerol (MGDG, >50% of the lipid content), digalactosyldiacylglycerol (DGDG), and sulfoquinovosyldiacylglycerol (SQDG), which is absent in Gloeobacterales (Rast et al. 2015). MGDG facilitates TM curvature (Bottier et al. 2007), while DGDG stabilizes thylakoid stacking via hydrogen bonding (Demé et al. 2014). The process by which TM are formed, and how their lipid ratios differ compared to those in the CM, has been the subject of four main hypotheses: membrane fusion, vesicular transport, direct lipid transport via soluble carriers (Jouhet et al. 2007; Rast et al. 2015) or self-assembly via a hexagonal-to-lamellar lipids phase transition (Guéguen and Maréchal 2022) (**Figure 2A**). Recent findings seem to support two of these hypotheses. First, the emergence of SQDG biosynthesis at the root of the Phycobacteria may have provided sufficient anionic lipids to enable the spontaneous formation of TM without requiring energy input (Guéguen and Maréchal 2022). Second, recent work by Tan et al. 2024 revealed that a duplication of the *PspA* gene in the ancestor of Phycobacteria, coupled with the acquisition of a C-terminal extension, gave rise to a vesicle-inducing protein, VIPP1, which supports the vesicular transport hypothesis (Tan et al. 2024). However, the role of VIPP1 in cyanobacteria remains to be fully understood. For instance, VIPP1 is critical for TM biogenesis in *Synechocystis* sp. PCC 6803 (Gao and Xu 2009), yet it appears to be non-essential in *Synechococcus* sp. PCC 7002 (Zhang et al. 2014). Additionally, other proteins found in cyanobacteria, such as CPSAR1 (Garcia et al. 2010), THF1 (Wu et al. 2011), and CPRabA5e (Karim et al. 2014), have been identified as essential for TM lipid biogenesis in *Arabidopsis thaliana*, although their role in cyanobacteria has never been investigated.

Unlike membrane structures, the transfer of transmembrane photosynthetic complexes has never been studied from an evolutionary perspective. To understand this transition, the biogenesis of these complexes—and specifically the proteins associated with it—provides crucial insights. The integration of LET complexes into TM during biogenesis is a highly regulated stepwise process involving numerous assembly factors (AF) that act as chaperones, notably for transporting PS subunits from CM to TM (**Figure 2B**) (Rast et al. 2015). The most studied element is PSII, whose reaction center (RC) biogenesis starts in CM with the precursor of D1, pD1. D1 transitions through the *pratA*-defined membrane (PDM), where D1 assembles with D2, leading to RC photoactivation (Klinkert et al. 2004; van de Meene et al. 2006; Rast et al. 2019). PDM, named after the *pratA* gene, also depends on *curT* (Rast et al. 2019) and *ancM* (Ostermeier et al. 2022), and serves as a contact point between TM and CM, playing a structural role in TM biogenesis (Rast et al. 2019). Once the PSII RC is assembled, it is first incorporated into TM with the CP43 and CP47 antenna complexes, followed by the addition of PSII subunits (Q, U, O, V) (Heinz et al. 2016). The complex then dimerizes and associates with the phycobilisome on the outer TM face (Heinz et al. 2016). Thirty-four AF coordinate this process in *Arabidopsis*, among which twelve are also known in cyanobacteria (**Table 1**). In contrast, the biogenesis of PSI is significantly less understood but nonetheless involves at least thirteen AF, of which only one—associated to PsbA subunits—has been detected in the CM. No evidence suggests that the PSI RC is assembled and activated outside the TM. Some subunits of the Cytb6f and ATPase complexes have also been detected in the CM, suggesting a stepwise assembly process, although only few corresponding AF have been reported to date (**Table 1**).

The complexity of the stepwise biogenetic process, and the number of coordinated AF involved in it, suggests the existence of evolutionary intermediate states stemming from the ancestral *Gloeobacter*-like state. Deciphering these intermediates involves inferring what the transfer sequence of the complexes was and what the evolutionary advantages of these states could have been. In addition to LET (**Figure 2C**), alternative electron flows (AEF) (**Figure 2D**) also operate in modern TM and notably compensate for the ATP/NADPH imbalance of OxyP.

Interestingly, except for the cyclic pathway involving the respiratory complex NDH-1, all AEF require fewer structural transfers (complexes or electron carriers) from the CM to TM compared to LET (**Figure 1D**), which involves six transfers. Thus, AEF may have become functional more quickly than LET during TM evolution. For this reason, I propose an evolutionary model for the origin of TM where intermediate states were primarily associated with AEF. Structurally, such an organism possessed a primordial TM (PriTM) dedicated to AEF, while maintaining a complete LET within its CM (**Figure 1B**). This PriTM would have been situated at the periphery of the cell due to the fact that the biogenetic process is parietal (Huokko et al. 2021). From an ultrastructural perspective, this cell would have resembled modern cyanobacteria, thus potentially uncoupling thylakoid structures from necessarily efficient OxyP in fossil cyanobacteria. For (pseudo)cyclic pathways, this implies that PriTM were initially associated with high ATP production. AEF linked to sulfide oxidation—anoxygenic photosynthesis (AnOxyP)—may have contributed to NADPH production in these PriTM. In both cases, PriTM would have contained at minimum an ATPase and PSI. Such intermediate states would have provided a key evolutionary advantage for the survival of early cyanobacteria.

The emergence of Phycobacteria is estimated to have occurred slightly before the GOE (Sánchez-Baracaldo et al. 2017; Fournier et al. 2021; Sánchez-Baracaldo et al. 2022) or during the early Proterozoic (Shih et al. 2017). This geological timespan enclosed the transition from an anoxic ocean to a post-GOE oxic ocean, but marked by euxinic episodes—i.e., anoxic and rich in sulfide—as evidenced by sulfur isotope ratios and the presence of $FeS_2$ in rocks (Scott et al. 2011). However, sulfide poses a significant challenge to OxyP due to its toxicity as a potent inhibitor of PSII (Cohen et al. 1986; Garcia-Pichel and Castenholz 1990; Miller and Bebout 2004), responsible for water splitting in OxyP. The presence of a PriTM dedicated to AEF, capable of producing ATP and potentially NADPH under euxinic conditions, would have conferred a major advantage. This could resolve the paradox of OxyP development in a sulfide-rich environment. The recent discovery of the sulfide quinone oxidoreductase (SQR) gene, associated with anoxygenic $H_2S$-related photosynthesis, in the *Thermostichales* (Tan et al. 2024)—the earliest-emerging order of Phycobacteria—further supports this hypothesis.

The alternative possibility of a transfer of PSII before PSI into PriTM does not appear to confer any significant evolutionary advantage without the rest of the LET chain and is therefore not favored in the present hypothesis. However, a late transfer of PSII would have made OxyP functional in TM and potentially triggered the GOE due to its radically increased efficiency. From a biogenesis perspective, the photoactivation of PSII outside TM, not observed for PSI, may be indicative of this sequential difference in the integration of PS complexes. From an evolutionary perspective, the apparition of PriTM occurred after the separation of Gloeobacterales but before the diversification of Phycobacteria. Future evolutionary studies on AF should help refining this hypothesis. AF known to be involved in the biogenesis of TM are listed in **Table 1**. Analyses of their distribution in cyanobacteria, their evolution, as well as the molecular dating of these AF, will in the future allow the determination of the transfer sequence.

In conclusion, the emergence of PSI-equipped PriTM dedicated to AEF facilitated ATP and potentially NADPH production, enabling survival in sulfide-rich environments of early earth. While these adaptations may not have directly supported long-term fitness, they represented a critical evolutionary step toward the development of the complex photosynthetic systems of modern cyanobacteria.

# Figures & Table

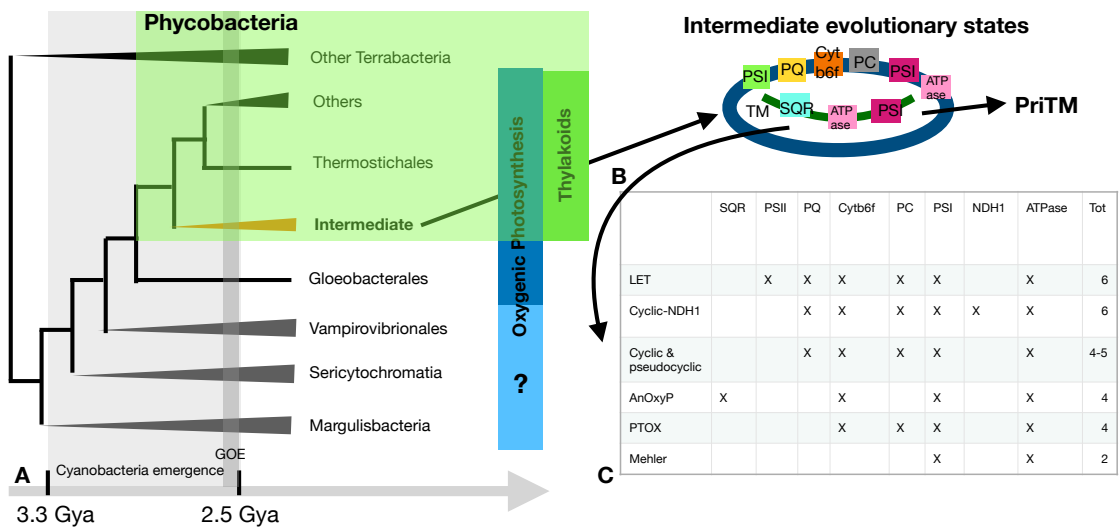

**Figure 2: Evolutionary Hypothesis for the Emergence of Thylakoid Membranes.**
**A. Evolutionary tree of cyanobacteria and their allies.** The current non-photosynthetic groups (gray triangles) and the two early branching groups important in the context of TM emergence, the Gloeobacterales and the Thermostichales, are represented. Phylogeny used from Gisriel et al. 2023. **B. Hypothetical evolutionary state.** Evolutionary state with a full LET in CM and a PSI-equipped PriTM dedicated to AEF. **C. Number of complexes used in AEF in PriTM**. The hypothesis involving the fewest transfers of complexes between CM and TM, and the most likely given the sulfide-rich environmental conditions of the Archaean era, points to alternative electron flows (AEF). The table summarizes the number of complexes involved in the different AEF.

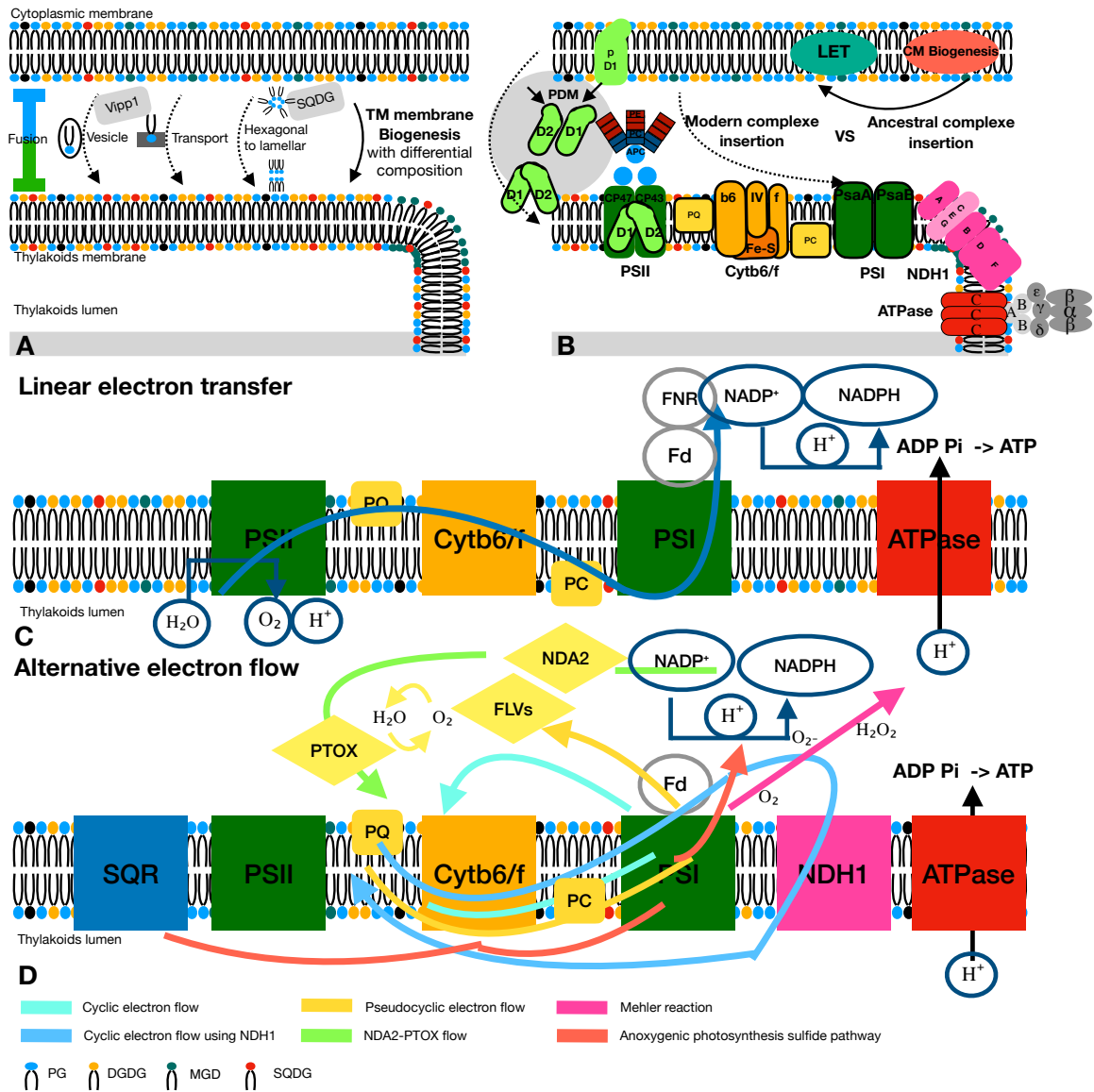

**Figure 1: Schematic view of thylakoid membrane emergence theory, biogenesis, and the roles of LET and AEF.**

**A. Four TM emergence theories and differential composition. CM** and **TM** are composed of the same lipidic components: **PG**, **MGDG**, **DGDG**, **SQDG** with proportions varying between the two membranes. Four scenarios have been proposed for the biogenesis of **TM** from **CM**. **B. Complex biogenesis.** In the primitive state, within **CM** of Gloeobacterales, complexes of the electron transport chains are located in specialized regions of **CM**, which also contain the biogenic regions. In the modern state, these complexes are found in **TM**, with the biogenesis of certain subunits starting in **CM** in a highly coordinated process; specifically, the entire reaction center (**RC**) of **PSII** is assembled in **PDM**, outside of **TM**. **C. Linear Electron Transfer.** **LET** is represented in blue. Light energy is captured by phycobilisomes, composed of **PE**, **PC** and **APC**, before being transmitted to the chlorophyll present in the **RC** of **PSII**. Electrons originating from the dissociation of $H_2O$ are transferred to **FNR**, passing through **PQ**, **Cytb6f**, **PC**, **PSI** and **Fd**. **D. Alternative electron flow.** The cyclic electron flow (light blue) reduces PSI components while bypassing PSII, giving electrons directly to PQ and creating ΔpH; **the cyclic pathway the NDH1 (blue)** complex to generate ΔpH; **the pseudo-cyclic flow (yellow)** transfers electrons to **FLV** to reduce $O_2$ to $H_2O$, with electrons being cyclically transferred back to PSI via **Cytb6f**, generating ΔpH; **the PTOX pathway (green)** reduces $O_2$ to $H_2O$ using electrons directly from NADPH via NADPH reductase; **the Mehler reaction (pink)** involves reduction of $O_2$ to $H_2O$, producing different reactive oxygen species (ROS) such as $O_2^-$ or $H_2O_2$. The Mehler reaction can also generate a ΔpH across thylakoid membranes, although the mechanism remains unclear; **the sulfide pathway, AnOxyP, (orange)** oxidize $H_2S$ using **SQR**, and electrons are transferred to **Cytb6f** before reaching PSI, generating ΔpH and NADPH. **PG** phosphatidylglycerol (blue), **MGDG** monogalactosyldiacylglycerol (green), **DGDG** digalactosyldiacylglycerol (yellow), **SQDG** sulfoquinovosyldiacylglycerol (Red). **APC** Allophycocyanin, **PC in blue**: Phycocyanin, **PE** Phycoerythrin, **PSII** Photosystem II, **PSI** Photosystem I, **Cytb6f** Cytochrome b6f complex, **PQ** plastoquinone, **PC in yellow** plastocyanin, **Fd** ferredoxin. **NDH1** NADH dehydrogenase complex 1, **PDM** PratA-defined membrane, **FNR** ferredoxin-NADP$^+$ reductase, **FLV** flavodiiron proteins, **SQR** sulfide quinone oxidoreductase, **PTOX** Plastid Terminal Oxidase. Alternative Electron flow. Modified from (Cohen et al. 1986; Blankenship 2010; Rast et al. 2015; Heinz et al. 2016; Rast et al. 2019; Huokko et al. 2021; Guéguen and Maréchal 2022; Eckardt et al. 2024).

| Complexes | Assembly factors |
|---|---|
| PSI (13) | **Alb3 (PDM, TM) ; VIPP1 (TM) ; Ycf3 (TM) ; Ycf4 (TM) ; Ycf37 (TM)** ; Y3IP1 (TM) ; PPD1 (TM) ; Psa2 (L) ; **RubA (TM) ; Hcf101 (S) ; CnfU (S)** ; APO1 (S); **Ycf51 (TM)** |
| PSII (34) | **ChlG (CM, PDM) ; CtpA (PDM, CM); CyanoP (CM)** ; HliA, HliB (PDM) ; **HliC, HliD (PDM)** ; Pittc (-) ; Pam68 (PDM, TM) ; **PratA (PDM) ; Psb27 (TM) ; Psb28 (TM)** ; Psb29 (-) ; Psb32 (TM) ; **Psb34 (-)** ; Psb35 (PDM) ; **SecY (PDM)** ; RubA (CM) ; Sll0408 (L) ; Sll0606 (CM) ; **Sll0933 (TM)** ; Slr0151 (TM) ; Slr0144 (-) ; Slr0286 (-) ; Slr0565 (CM) ; Slr1761 (L) ; Slr2013 (TM, PDM) ; **Ycf39 (PDM) ; Ycf48 (PDM, TM) ; YidC (PDM, TM)** ; LP2/3 (TM) ; **LPA19 (L)** ; PsbP (PDM) ; PsbN (PDM,CM) |
| Cytb6f (4) | HCF164 (-); trxm134 (-) ; NTA1 (-) ; DEIP1 (-) |

**Table 1: List of 51 known assembly factors**.
AF in bold are found in cyanobacteria, all AF at the exception of Ycf51 are also found in eukaryotic chloroplasts. Localization of the AF are indicated in parentheses. Modified from (Chi et al. 2012; Rast et al. 2015; Yang et al. 2015; Heinz et al. 2016; Johnson and Pakrasi 2022; Sandoval-Ibáñez et al. 2022; Li et al. 2023; Chen et al. 2024; Dai et al. 2024).

# Funding

This work was supported by a research grant (PDR T.0018.24 OR-OX-PHOT-IN-CYN) financed by the Belgian National Fund for Scientific Research (FNRS).

# Conflict of Interest

The author declares no competing interest.